# High Pressure Induced Binding Between Linear Carbon Chains and Nanotubes

Gustavo Brunetto[2], Nádia F. Andrade[1], Douglas S. Galvão[2], and Antônio G. Souza Filho[1]
[1]Physics Department, Federal University of Ceará, 60440-900, Fortaleza, Ceará, Brazil
[2]Applied Physics Department, State University of Campinas, 13083-970 Campinas, São Paulo, Brazil.

## ABSTRACT

Recent studies of single-walled carbon nanotubes (CNTs) in aqueous media have showed that water can significantly affect the tube mechanical properties. CNTs under hydrostatic compression can preserve their elastic properties up to large pressure values, while exhibiting exceptional resistance to mechanical loadings. It was experimentally observed that CNTs with encapsulated linear carbon chains (LCCs), when subjected to high hydrostatic pressure values, present irreversible red shifts in some of their vibrational frequencies. In order to address the cause of this phenomenon, we have carried out fully atomistic reactive (ReaxFF) molecular dynamics (MD) simulations for model structures mimicking the experimental conditions. We have considered the cases of finite and infinite (cyclic boundary conditions) CNTs filled with LCCs (LCC@CNTs) of different lengths (from 9 up to 40 atoms). Our results show that increasing the hydrostatic pressure causes the CNT to be deformed in an inhomogeneous way due to the LCC presence. The LCC/CNT interface regions exhibit convex curvatures, which results in more reactive sites, thus favoring the formation of covalent chemical bonds between the chain and the nanotube. This process is irreversible with the newly formed bonds continuing to exist even after releasing the external pressure and causing an irreversibly red shift in the chain vibrational modes from 1850 to 1500 cm$^{-1}$.

## INTRODUCTION

Recently, studies on single-walled CNTs behavior on aqueous media were reported [1,2]. These studies showed that water can significantly affect the tube mechanical, but the compressed structures still preserve their elastic strength properties and an exceptional resistance to mechanical loading. These studies are suggestive that CNTs can be good platforms for applications in nanofluidic devices.

It was experimentally observed [3] that when subjected to high pressures, linear carbon chains (LCCs) inside carbon nanotubes (CNT) present irreversible red shifts for some of their vibrational frequencies due to the LCC coalescences induced by the external pressure. In this work we have investigated, through fully atomistic molecular dynamics (MD) simulations, the behavior of single LCC confined inside CNT and subject to extreme high external hydrostatic pressures (up to 10 times higher than used to merge two LCC). Our results show that depending on the applied pressure values, the induced structural deformations can lead to the formation of covalent chemical bonds between the LCC and the tube walls. These newly formed bonds remain stable even when the applied pressures are removed. Similarly to the case where two LCC are merged, some vibrational frequencies are permanently red shifted.

## THEORY

In order to address the structural aspects of LCCs inside CNTs subject to extremely high pressures, we have carried out MD simulations using the reactive force-field ReaxFF, which is a Bond Energy Bond Order (BEBO) method [4,5]. ReaxFF can describe bond breaking and bond formation during chemical reactions. Bond orders are calculated from interatomic distances, as a sum of σ, π and π-π terms [5].

Similar to empirical nonreactive force fields, the BEBO system energy is composed of a sum of different components, such as; bonding, bending angles, dihedrals, Coulomb [4,5]. Charge distributions [6] are calculated based on geometry and connectivity using the electronegativity equalization method (EEM) [7]. Numerical simulations were carried out using LAMMPS code [8] with the ReaxFF method there implemented [9].

We have considered the cases of finite CNTs with capped closed ends, as well as, infinite ones (cyclic boundary conditions along the tube axial direction) with encapsulated LCCs of different lengths (9 and 40 carbon atoms for the finite and infinite case, respectively). For both cases we used (5,5) CNT, which has a diameter of 6.8 Å. Tubes with lengths of 30.8 and 73.6 Å for the finite and periodic cases, respectively, were considered. These tubes were chosen to be representative and to mimic the experimental conditions [3].

For the finite case we used a simulation box with dimensions of 29×29×54 Å$^3$. The frontier edges are composed of rigid reflective walls, which deflect the atoms when they reach the box walls (they are elastically scattered). The space between the CNT and the box walls is then filled with water molecules, as shown Figure 1-a. For the infinite case the box dimensions were of 32.0×32.0×73.6 Å$^3$. A typical cross-section view of the unit cell for the periodic system is shown in Figure 1-b with the LCC composed of 40 atoms (highlighted in yellow).

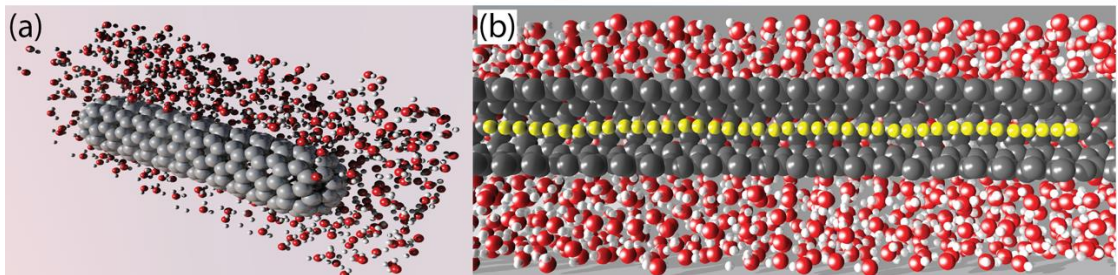

**Figure 1**. (a) Capped carbon nanotube immersed into a water environment. A carbon linear chain (not visible) composed of 9 carbon atoms is inside the tube; (b) Cross-section of the box simulation. A linear chain composed of 40 carbon atoms (indicated in yellow) is placed inside a periodic CNT immersed into a water environment.

In order to increase the external pressure experienced by the CNT, the *x* and *y* box dimensions were continuously decreased with a constant rate of $v_{wall}=2\times10^{-5}$ Å/ps. The box dimension along the tube axial direction (*z*-dimension) was kept constant for all cases. The simulations were carried out using a canonical NVT ensemble with constant target temperature of T=300K.

The vibrational density of states ($\Phi(\omega)$) of the chains were obtained by the Fourier transform of the velocity auto-correlation function (VAF) (Z(t)) (Equations 1 and 2, respectively).

$$\Phi(\omega) = \left[\frac{1}{\sqrt{2}} \int_{-\infty}^{\infty} dt \, e^{i\omega t} Z(t)\right]^2 \quad (1)$$

$$Z(t) = \frac{\langle \mathbf{v}(0) \cdot \mathbf{v}(t) \rangle}{\langle \mathbf{v}(0) \cdot \mathbf{v}(0) \rangle} \quad (2)$$

The corresponding vibrational spectra were obtained using equations (1) and (2). We considered the cases of bonded (as a consequence of the externally applied pressure) and non-bonded (before applying pressure) LCCs. To obtain the simulated spectra, first the system composed of the CNT and LCC is thermalized using a NVT ensemble (T=300K) during 20 ps. After thermalization the system is placed into an adiabatic box and let to evolve with constant total energy (NVE ensemble) during 500 ps. During the second step only the velocities of the atoms belonging to the LCC are recorded from which the velocity autocorrelation function Z(t) is computed.

**DISCUSSION**

The temporal evolution of the pressure values for the periodic CNT is shown in Figure 2. Besides the expected tube deformation due to the hydrostatic pressure action, after the system went a critical stage, when the distance between LCC and CNT is decreased, a covalent chemical bond between is formed. The red point in Figure 2 indicates the instant where this critical pressure is achieved and reaction occurs.

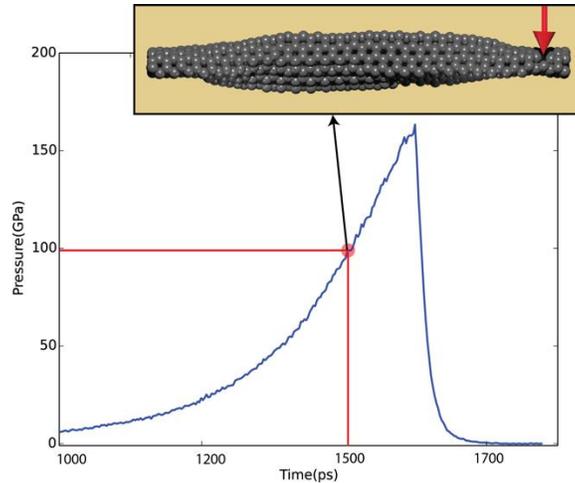

**Figure 2**. Pressure temporal evolution values during hydrostatic compression for the case of a periodic CNT. The highlighted red point indicates the instant where the confined LCC is covalently bonded to the CNT wall. The inset shows the inhomogeneous tube deformation, the center tube part where the LCC is located is less deformed than its extremities. The tube regions that are not into direct contact with the LCC atoms (indicated by a red arrow) are more flexible and, therefore, more deformable.

Through the reactive MD simulations was possible to follow and identify the structural changes in LCC/CNT system as a function of the externally applied hydrostatic pressures. The first observed effect was that as the pressure is increased, the tube walls start to deform and are pulled against the LCC atoms. The parts of the tube walls that are into direct contact with the LCC become less flexible (as the LCC atoms structurally resist to the 'mechanical squeezing'). The remaining regions are more flexible and easily deformable under pressure, as illustrated in the inset to Figure 2. The net result is a CNT with its geometry deformed in a non-homogeneous way.

The chemical reaction can be also analyzed from an energy viewpoint. The increase of the pressure decreases the distance between the LCC and the tube walls, thus resulting in an increase in the potential energy (Figure 3). Energy continuously increases up to the instant t=1492 ps, which corresponds to a pressure of 109 GPa (indicated by the red box in Figure 3). The MD analysis showed that the potential energy drop off by $\Delta E = 37$ Kcal/mol, indicating the instant where a covalent bond between LCC and the tube is formed. This is confirmed by measuring the distance between the atoms involved in the reaction (red curve in Figure 4-a). The reaction occurs between one of the atoms belonging to CNT wall and one of the LCC ends, labeled by C1 and C2 in Figure 4-b. The created new bond constitutes a stable configuration, since it continues to exist even for higher pressures (corresponding to the increase in energy between 1492 and 1610 ps in Figure 3). More importantly, the chain remains attached to the CNT even when the external pressure is released, indicating that the process is irreversible, which is consistent with the available experimental data.

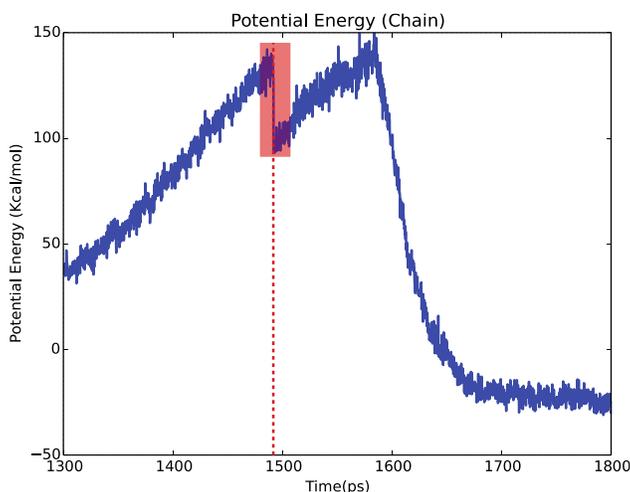

**Figure 3**. Chain potential energy during the procedure of compression (from t=1300 to t=1600 ps) and decompression (from t=1600 to t=1800 ps). The red line indicates that at in the instant 1492 ps, the pressure of 109 GPa is enough to induce the reaction between the LCC and CNT, causing an instantaneous drop of the potential energy.

From the MD simulations it was also possible to observe that this covalent bonding involves one atom from the LCC ends and an atom from tube walls. This pattern is the same for the finite and infinite cases, indicating that the capped tube regions are not the bonding preferential regions.

A geometrical analysis at the instant where bond is formed allows us to understand why the bond formation occurs. The atoms belonging to chain ends are in a dangling bond-like state. The CNT compression creates regions with different levels of deformations at the LCC/CNT interface, these regions having higher curvatures than the isolated CNT. In general, higher curvatures means higher chemical reactivity [10-12], which lowers the potential barrier to make a bond between the LCC and CNT wall.

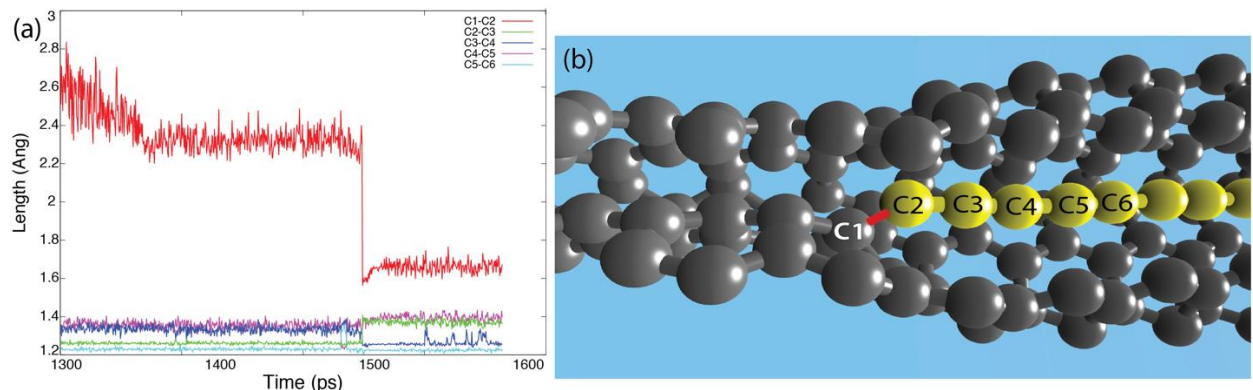

**Figure 4**. (a) Bond-length values between the carbon atoms involved in the reaction (labels are showed in (b)). When the pressure reaches the critical value to allow bond formation, the distance between atoms C1 and C2 drops from 2.3 to 1.6 Å, characterizing the instant reaction; (b) MD snapshot showing the LCC (in yellow) covalently bonded to CNT (in gray) during the hydrostatic compression. In order to facilitate the visualization of these processes the water atoms and some of CNT atoms were made transparent.

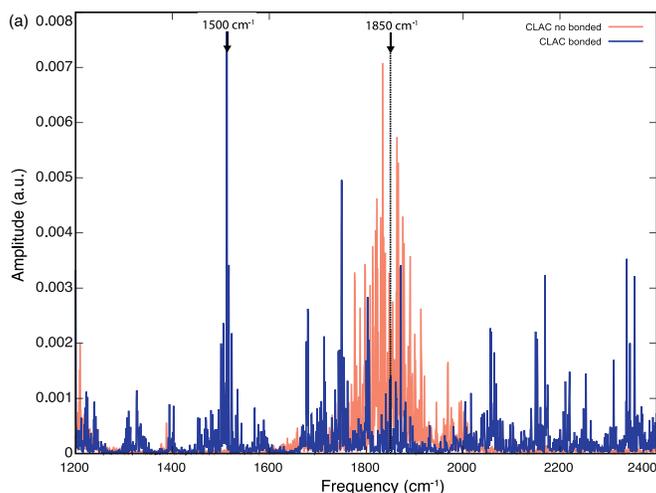

**Figure 5**. Vibrational frequency for the system composed of a capped CNT with a 9 atoms encapsulated LCC showing the spectra differences before (in red) and after (in blue) the hydrostatic compression.

In order to understand how the reaction between LCC and CNT affect the chain vibrational spectra, we simulated these spectra through the velocity auto-correlation method (as described in the methodology section) for capped finite and periodically infinite CNTs. As we can see from Figure 5 (results for capped CNT), the simulated spectrum exhibits the expected carbon chain characteristic peak at 1850 cm$^{-1}$ [14,15]. When a bond between the chain and CNT is created, the spectrum is red shifted to 1750 cm$^{-1}$ and a new strong peak appears at 1500 cm$^{-1}$. The results are similar for an infinite CNT with an encapsulated LCC composed of 40 atoms.

# CONCLUSIONS

We have investigated, through fully atomistic reactive molecular dynamics simulations (ReaxFF), the structural aspects of linear carbon chains (LCCs) encapsulated into carbon nanotubes (CNT) (LCC@CNT). The LCC@CNT systems were immersed into a water environment and subjected to externally applied pressures. We have considered the cases of finite and infinite (cyclic boundary conditions) CNTs with LCCs of different lengths (composed of 9 and 40 atoms). Very similar results were obtained for both cases. Our results show that during the compression CNT deforms in an inhomogeneous way due to the presence/absence of the LCC. The interface regions LCC/CNT exhibit a convex curvature, which is reflected by a chemically reactive environment that favors covalent bond formation between LCC and CNT. This process showed to be irreversible with the newly formed covalent bonds continuing to exist even when the applied external pressure is released. The new bonds cause a peak red shift from 1850 to 1750 $cm^{-1}$ in the vibrational spectra. Also, a new peak appears at 1500 $cm^{-1}$, which is consistent with the newly formed chemical bond and with the available experimental data.

# ACKNOWLEDGMENTS

This work was supported in part by the Brazilian Agencies CAPES, CNPq and FAPESP. The authors thank the Center for Computational Engineering and Sciences at Unicamp for financial support through the FAPESP/CEPID Grant # 2013/08293-7.